\documentclass[twocolumn,showpacs,preprintnumbers,amsmath,amssymb,nofootinbib]{revtex4-2}

\usepackage{graphicx}
\usepackage{dcolumn}
\usepackage{bm}

\usepackage{epstopdf}%
\usepackage{amsmath}%
\usepackage{amsfonts}%
\usepackage{amssymb}
\usepackage{xcolor}
\usepackage{tikz}
\usetikzlibrary{positioning}
\definecolor{Mycolor2}{HTML}{FF00FF}

\begin{document}

\title{Topological edge states of the hexagonal linear chain}

\author{M. Ni\c t\u a}
\address{National Institute of Materials Physics, Atomistilor 405A,
Magurele 077125, Romania}


\begin{abstract}

We study the eigenspectrum properties of a one-dimensional molecular chain composed of hexagonal 
unit cells.
The system features two alternating hopping parameters, resulting in a rich energy spectrum with both dispersive and flat bands.
By analyzing the model under periodic and open boundary conditions, 
we identify two insulating phases separated by a gap-closing transition controlled by the ratio of hopping amplitudes.
In the topological phase, realized when the hopping ratio falls below a critical value, 
edge states emerge that are exponentially localized at the boundaries of finite chains.

\end{abstract}

\maketitle

\section{Introduction}

The Hückel discrete Hamiltonian of a linear chain with alternating hopping energies 
was first introduced by Lennard Jones \cite{lennard1937} to describe orbital energies in certain hydrocarbon chains. 
Since 1980 \cite{su1980}, it has become known as the Su-Schrieffer-Heeger (SSH) model, 
representing the simplest one-dimensional (1D) system that exhibits topological properties and edge states \cite{batra2020, ostahie2021}. 
The modulation of alternating hopping amplitudes drives a transition between two insulating phases and, 
under specific conditions, gives rise to localized edge modes.

Beyond the SSH model, various quasi-1D systems have been shown to support topological behavior and edge states, 
such as extended SSH chains and trimer lattices 
\cite{alvarez2019, anastasiadis2022, dasallas2024, verma2024, guo2025, wielian2025, mohammad2025}. 
These systems are also of interest for potential applications in electronics, photonics, and acoustics 
\cite{antonin2021, lan2022, wang2022, gabriel2022, guo2024, alexis2025}.


In this article, we demonstrate the emergence of topological edge states
in a molecular chain consisting of hexagonal atomic units connected in series.
The hopping amplitudes along the chain alternate between two values, $t_1$ and $t_2$.
Unlike the conventional SSH model, which features a two-site unit cell,
the system studied here has a six-atom unit cell, {while remaining in the same symmetry class as the standard SSH model, 
despite its richer internal structure.
This increased structural complexity leads to a richer energy spectrum: 
the larger unit cell gives rise to multiple bands, while the hexagonal geometry introduces interference effects that can generate flat bands.
Moreover, the presence of two distinct tunneling paths across each hexagon modifies the effective coupling along the chain, 
causing the edge states to emerge at lower values of hopping energies ratio than in the standard SSH model.}

We show that, for periodic boundary conditions, 
the system displays an energy gap that closes when the hopping amplitudes satisfy $\sqrt{2} t_1 = t_2$. 
In 1D topological models, such a gap closing typically signals a topological phase transition, 
accompanied by the appearance of edge states, as predicted by the bulk-boundary correspondence principle. 
We investigate whether a similar transition occurs in the hexagonal chain and demonstrate that, 
for hopping ratios $r = \sqrt{2} t_1 / t_2$ below a critical value $r_c$, 
two eigenstates become exponentially localized at the chain's edges, forming topological edge states. 
This behavior disappears for $r > r_c$, and we further analyze how the critical ratio $r_c$ depends on system size.


\section{Hexagonal chain}

\begin{figure}[h]
\centering
\includegraphics[scale=0.7, trim=0 20 0 20]{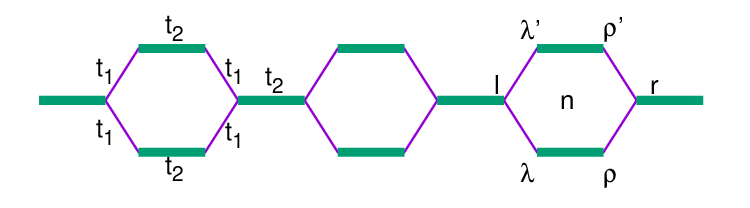}
\caption{
A nanowire formed by artificial benzene molecules coupled in series in the \emph{para} position.
Each unit cell consists of six atomic sites labeled as $s = l, r, \lambda, \lambda', \rho, \rho'$.
The cells are indexed by $n = 1, \cdots, N$.
}
\label{lant1}
\end{figure}
We consider a linear hexagonal chain as shown in Fig.\,\ref{lant1}, consisting of $N$ hexagonal unit cells. 
Each unit cell contains six atomic orbitals denoted by $s = l, r, \lambda, \lambda', \rho, \rho'$.

The system is described by the following single-particle Hamiltonian:
\begin{eqnarray} \label{hlant}
H=\sum_{n=1}^N   t_1(|nl\rangle \langle n\lambda |+     |nl\rangle \langle n\lambda' |      
                   +|n\rho \rangle \langle nr |+|n\rho' \rangle \langle nr |) \nonumber \\
                 t_2(|n\lambda \rangle \langle n\rho |+     |n\lambda' \rangle \langle n\rho' |      
                   +|nr \rangle \langle n+1 l |)    +h.c., ~~ ~~  
\end{eqnarray}
where $t_1$ and $t_2$ are alternating hopping amplitudes along the chain.
{The vectors $|ns\rangle$ with $n=1,\cdots, N$ and $s=l,\cdots,\rho'$ 
form a localized orbital basis, where each state describes a particle occupying orbital 
$s$ in unit cell $n$}.
An experimental realization of this system is possible in various physical platforms 
such as quantum dot arrays with adjustable tunneling rates \cite{tamura2002, volk2019, beatriz2019, simmons2022}
or photonic structures \cite{lan2022, alexis2025}.

Before proceeding with the explicit calculation of the eigenstate properties of this system, 
we note that certain symmetry properties of the Hamiltonian can be exploited to infer useful characteristics of the solutions. 
For this, one considers the general solution of the Schrödinger equation:
\begin{equation} \label{seq}
H|\Psi\rangle = E|\Psi\rangle. 
\end{equation}

The Hamiltonian $H$ anticommutes with the chiral symmetry operator $\Gamma$ defined as \cite{ostahie2021}:
\begin{equation} \label{ch1}
\Gamma =   
 \sum_{n, s = r, \lambda, \lambda'} |ns\rangle \langle ns|
 -\sum_{n, s =l, \rho, \rho'} |ns\rangle \langle ns|. 
\end{equation}
This anticommutation implies electron-hole symmetry in the energy spectrum. 
If $|\Psi\rangle$ is an eigenstate of $H$ with energy $E$ from Eq.\,\ref{seq}, 
then $\Gamma|\Psi\rangle$ is also an eigenstate with opposite energy:
\begin{equation} \label{ch2}
H(\Gamma|\Psi\rangle) = -E(\Gamma|\Psi\rangle). 
\end{equation}

The hexagonal chain also exhibits parity symmetry, expressed as $[H, P] = 0$, where the parity operator $P$ is given by:
\begin{eqnarray} \nonumber
P &=& \sum_{n} \left( |n\lambda\rangle \langle n\lambda'| + |n\rho\rangle \langle n\rho'| \right) \\ \label{par}
&& + \frac{1}{2} \sum_{n, ~s = l, r} |ns\rangle \langle ns| + \text{h.c.}. 
\end{eqnarray}

This symmetry allows the Schrödinger equation \eqref{seq} to be solved by restricting the analysis to eigenstates of definite parity, 
i.e., states that satisfy $P|\Psi\rangle = \pm|\Psi\rangle$. 
In doing so, only symmetric or antisymmetric wavefunctions are considered as candidates for eigenstates of $H$. 
This significantly reduces the size of the solution space and simplifies the analysis of the system’s spectrum.

\subsection{Eigenstates properties of the circular chain}

For the circular chain, periodic boundary conditions are imposed: $|N+1,l\rangle = |1,l\rangle$. 
In this case, we work in $k$-space representation, defining new vectors $|k,s\rangle$ such that
\begin{eqnarray}\label{ns}
|ns\rangle=\frac{1}{\sqrt N} \sum_k e^{-ikn}  |ks\rangle,~
\end{eqnarray}
with wave number $k=\frac{2\pi}{N}j$ and integer $j$ chosen such that $|n,s\rangle=|n+N,s\rangle$.
$N$ consecutive values of the integer $j$ are chosen.

The wave number $k$ is a good quantum number, and in the new basis $|k,s\rangle$,
the Hamiltonian becomes block-diagonal, $H=\sum_k H_k$, where each $H_k$ acts in a six-dimensional vector space:
\begin{eqnarray} \nonumber
	H_k=&& t_1 (| k\lambda \rangle\langle kl | + | k\lambda' \rangle\langle kl |+
	         | k\rho \rangle \langle kr |+ | k\rho' \rangle \langle kr |) ~~~~\\  \label{hka}
	    && + t_2 (  | k\lambda \rangle \langle k \rho |+ | k\lambda' \rangle \langle k \rho' | +
	         e^{ik} |kr \rangle \langle kl |) + h.c..
\end{eqnarray}

The Schrödinger equation (\ref{seq}) becomes:
\begin{eqnarray}\label{hkek}
H_k|\Psi_k \rangle = E_k |\Psi_k \rangle,
\end{eqnarray}
with eigenvectors expressed as:
\begin{eqnarray}\label{hkek1}
\Psi_k=\frac{1}{C}\sum_{n,s} \Psi_k(n,s)|n,s\rangle.~
\end{eqnarray}
These are obtained by solving the eigenvalue problem for $H_k$ given in Eq.\,\ref{hka}. 

We first compute the symmetric eigenstates of $H_k$ with parity $P=1$. The calculation method is shown in the
Appendix\,A and we find:
\begin{eqnarray} \nonumber
|\Psi_k ^ {\pm} \rangle = &&|k r \rangle + e^{- i \phi_k} |k l \rangle \\ \label{vk}
               && \pm \frac{1}{\sqrt 2} e^{-i k/2} \left( |k \lambda' \rangle +|k \lambda \rangle \right)  \\ \nonumber
               && \pm \frac{1}{\sqrt 2} e^{ i k/2 - i \phi_k} (|k \rho' \rangle +|k \rho \rangle ).           
\end{eqnarray}
The angle $\phi_k$ encodes the relative phase between the two partite subspaces 
defined by the chiral symmetry operator $\Gamma$ in Eq.\,\ref{ch1}.
By substituting these vectors into Eq.\,\ref{hkek}, we obtain the corresponding energy relation:
\begin{eqnarray}\label{ek}
E_k^{\pm} e^{i\phi_k^{\pm}} = t_2e^{ik} {\pm} \sqrt{2}{t_1}e^{i\frac{k}{2}}. 
\end{eqnarray}

The eigenvectors $|\Psi_k ^ {\pm}\rangle$ must be expressed in the $|n,s\rangle$ basis as in Eq.\,\ref{hkek1}.
By using Eq.\,\ref{ns}, the real-space components $\Psi_k(n,s)$ are found:
\begin{eqnarray}\label{psil}
\Psi_k^{\pm}(nl)&&=e^{ikn-i\phi_k},   \\ 
\label{psir}
\Psi_k^{\pm}(nr)&&=e^{ikn},  \\ 
\label{psilm}
\Psi_k^{\pm}(n\lambda )=\Psi_k^{\pm}(n\lambda' )&&=\pm\frac{1}{\sqrt 2}e^{ik\left(n-\frac{1}{2}\right)},\\ 
\label{psiro}
\Psi_k^{\pm}(n\rho )   = \Psi_k^{\pm}(n\rho' )  &&=\pm\frac{1}{\sqrt 2}e^{ik\left(n+\frac{1}{2}\right)-i\phi_k^{\pm}}, 
\end{eqnarray}
with normalization constant $C=\sqrt{4N}$.
An alternative method for calculating these eigenvectors may start with the Green’s function techniques, 
as it is used in 
\cite{malysheva2008, onipko2008, malysheva2024} for graphene type lattices.

From the energy equation (\ref{ek}), one can extract both the dispersion relation of $E_k$ and the value of 
the corresponding angle $\phi_k$.
The first two energy bands with positive energies, $E_k^{1,2} \equiv E_k^{\pm}$, are given by:
\begin{eqnarray}\label{ek2}
E_k^{1,2}  =\sqrt{t_2^2 + 2t_1^2 \pm 2\sqrt{2}t_1t_2\cos\frac{k}{2}}, 
\end{eqnarray}
and the angle $\phi_k$ (defined modulo $\pi$) satisfies:
\begin{eqnarray}\label{phik}
\tan \phi_k^{1,2}=\frac{t_2 \sin k {\pm} \sqrt2t_1 \sin\frac{k}{2} }{t_2 \cos k \pm \sqrt{2}t_1 \cos\frac{k}{2} }, 
\end{eqnarray}
ensuring that the resulting energies are positive.

The next two bands have negative energies and are obtained by the substitution $\phi_k \to \phi_k+\pi$ in Eq.\,\ref{ek}, 
resulting in $E_k^{3,4} = -E_k^{1,2}$ and $\Psi_k^{3,4}(n,s) = -\Psi_k^{1,2}(n,s)$ for $s = l, \rho, \rho'$.
This behavior reflects the particle-hole symmetry of the bipartite lattice, as expressed in Eqs.\,\ref{seq} and \ref{ch2}.


The last two bands $E^{5,6}$ are flat, with energies:
\begin{eqnarray}\label{eflat}
E^{5,6}=\pm t_2, 
\end{eqnarray}
and degeneracy $N$.
The degenerate eigenstates $|\Psi_j^5\rangle$ corresponding to energy $E^5=t_2$, with $j = 1, \cdots, N$, 
have the components:
\begin{eqnarray}\label{fb1}
\Psi_j^5(nl)&&=\Psi_j^5(nr)=0~\\ 
\label{fb2}
\Psi_j^5(n\lambda )&&=-\Psi_j^5(n\lambda' )=\frac{1}{C}\delta_{nj}~\\ 
\label{fb3}
\Psi_j^5(n\rho)&&=-\Psi_j^5(n\rho')=\frac{1}{C}\delta_{nj}, 
\end{eqnarray}
with normalization constant $C=2$.

\begin{figure}[h!]
    \centering

\begin{tikzpicture}[scale=0.55, every node/.style={font=\normalsize}]
    
    \node (1) at (0,0) [circle,draw,minimum size=4mm,inner sep=0pt,label=above:{0}] {l};
    \node (2) at (1,1.3) [circle,draw,minimum size=4mm,inner sep=0pt, label=above:{$-\frac{1}{2}$}] {$\lambda'$};
    \node (3) at (3,1.3) [circle,draw,minimum size=4mm,inner sep=0pt, label=above:{$-\frac{1}{2}$}] {$\rho'$};
    \node (2p) at (1,-1.3)  [circle,draw,minimum size=4mm,inner sep=0pt, label=below:{$\frac{1}{2}$}]{$\lambda$};
    \node (3p) at (3,-1.3) [circle,draw,minimum size=4mm,inner sep=0pt, label=below:{$\frac{1}{2}$}] {$\rho$};
    \node (4) at (4,0)  [circle,draw,minimum size=4mm,inner sep=0pt, label=above:{0}]{r};
    \node at (2,0) {$n=j$};

    \draw (1) -- (2) ;   
    \draw (1) -- (2p) ;
    \draw (3) -- (4) ;
    \draw (3p) -- (4)  ;

    \draw[line width=2.2pt, green!50!black] (2) -- (3)  ; 

     \draw[line width=2.2pt, green!50!black] (2p)--(3p);
       
    \draw[dotted, line width=2.2pt, green!50!black] (4) -- ++(1.5,0);
    \draw[dotted, line width=2.2pt, green!50!black] (1) -- ++(-1.5,0);


\end{tikzpicture}
\caption{{The localization of the flat band states $\Psi_j^{5}$ on the atomic sites of the $j^{th}$ hexagonal cell.
Following the opposite amplitudes on the upper and lower arms of the hexagon, 
the state cannot propagate along the chain as a result of destructive quantum interference.}}
\label{figfb}
\end{figure}
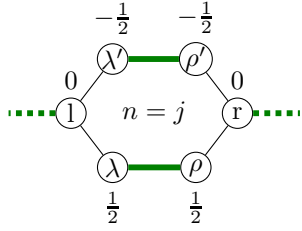

{A generic flat-band state $|\Psi_j^5\rangle$ is sketched in Fig.~\ref{figfb}. 
This state is constructed to be localized on the $j^{th}$
 unit cell of the nanowire, such that the amplitudes on the upper sites have opposite signs compared to those on the lower sites. 
 As a result, these states do not propagate along the chain due to destructive quantum interference, 
 as commonly occurs in flat-band systems \cite{nita2013, leykam2018, thomas2018, zeng2025}.}
They are antisymmetric states with $P = -1$ for operator in Eq.\,\ref{par}
and satisfy the equation $H|\Psi_j^5\rangle = t_2|\Psi_j^5\rangle$.
The remaining flat-band eigenstates follow from chiral symmetry (\ref{ch1}) 
via $\Psi_j^{6}(n,s) = - \Psi_j^{5}(n,s)$ for $s=\rho,\rho'$ and $E^6=-E^5$.

\begin{figure}[h]
\centering
\includegraphics[width=\columnwidth, trim=0 20 30 60]{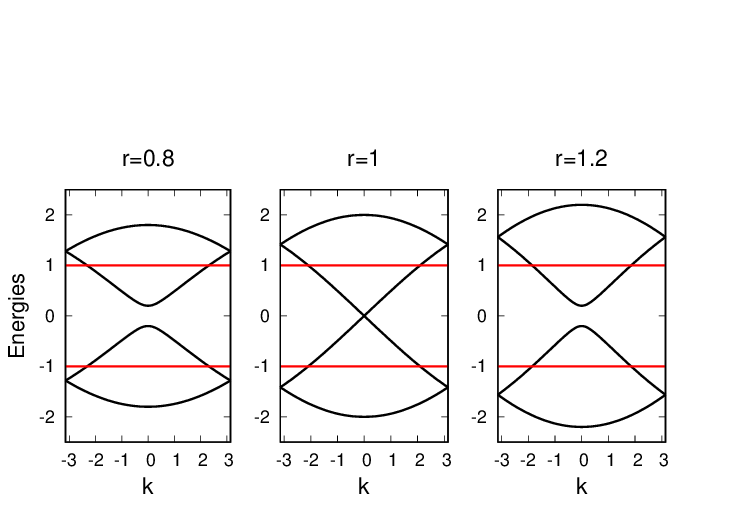}
\caption{The energy bands of the periodic chain versus wave number $k$.}
\label{energii}
\end{figure}

From the dispersion relation of the $E_k^{-}$ branch in Eq.\,\ref{ek}, 
we note that $E_k^{-} = 0$ occurs at $k=0$ when $t_2 = \sqrt{2} t_1$.
This motivates the introduction of the dimensionless parameter
\begin{eqnarray}\label{ratio}
r=\frac{\sqrt2 t_1}{t_2}, 
\end{eqnarray}
which identifies the closing of the energy gap at the critical value $r=1$.
This behavior is illustrated in Figure\,\ref{energii}, which shows the evolution of the energy bands as a function of $k$ for three distinct regimes: $r<1$, $r=1$, and $r>1$.
{The system exhibits two insulating phases for $r<1$ and $r>1$, separated by a gap closing at $r_c = 1$.
To reveal the topological nature of this transition, we first calculate the winding number of the chiral Hamiltonian 
$H_k$. This indicates the presence of topological edge states in the nontrivial phase, in agreement with the bulk-boundary correspondence.}

%



\subsection{Winding number of $H_k$}

{To quantify the topological properties, we write the Hamiltonian $H_k$ in matrix form.  
Due to its chiral symmetry (\ref{ch1}), the Hamiltonian assumes a block off-diagonal structure}:
\begin{eqnarray}
H_k=
\begin{bmatrix}
0 & D_k \\
D_k^* & 0
\end{bmatrix}.
\end{eqnarray}
The matrix $D_k$ is given by:
\begin{eqnarray}
D_k=
\begin{bmatrix}
t_2e^{ik} & t_1 & t_1 \\
t_1 &   t_2 & 0 \\
t_1  & 0  & t_2
\end{bmatrix}
\end{eqnarray}
and contains the elements between the two sublattices defined by the chirality operator (\ref{ch1}).

The topological properties of the system are encoded in the complex function behaviour $f(e^{ik}) = \det D_k$:
\begin{eqnarray}
\det D_k = t_2^3 \left( e^{i k} - \frac{2 t_1^2}{t_2^2} \right).
\end{eqnarray}
Zero-energy states arise from the condition $\det D_k = 0$, which occurs when $k = \pi$ and $\frac{2 t_1^2}{t_2^2} = 1$.
Under these circumstances, as shown in Ref.\,\cite{chen2024}, 
the topological phase is characterized by the winding number of $f(z)$, where $z = e^{ik}$ and $k \in [0, 2\pi]$.
The corresponding topological invariant is calculated:
\begin{eqnarray}
\nu = \frac{1}{2\pi i} \int_{BZ} \frac{\partial_k \det D_k}{\det D_k}  dk = \begin{cases}
1, & \text{if } r^2 < 1 \\
0, & \text{if } r^2 > 1
\end{cases}
\end{eqnarray}

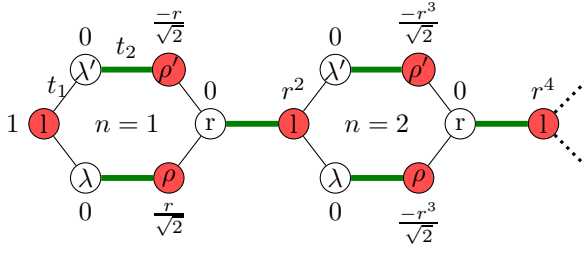
\begin{figure}[h!]
    \centering

\begin{tikzpicture}[scale=0.55, every node/.style={font=\normalsize}]
    
    \node (1) at (0,0) [circle,draw,minimum size=4mm,inner sep=0pt, fill=red!70, label=left:{1}] {l};
    \node (2) at (1,1.3) [circle,draw,minimum size=4mm,inner sep=0pt, label=above:{0}] {$\lambda'$};
    \node (3) at (3,1.3) [circle,draw,minimum size=4mm,inner sep=0pt,fill=red!70, label=above:{$\frac{-r}{\sqrt 2}$}] {$\rho'$};
    \node (2p) at (1,-1.3)  [circle,draw,minimum size=4mm,inner sep=0pt, label=below:{0}]{$\lambda$};
    \node (3p) at (3,-1.3) [circle,draw,minimum size=4mm,inner sep=0pt,fill=red!70, label=below:{$\frac{r}{\sqrt 2}$}] {$\rho$};
    \node (4) at (4,0)  [circle,draw,minimum size=4mm,inner sep=0pt,label=above:{0} ]{r};
    \node at (2,0) {$n=1$};
    \node at (8,0) {$n=2$};

    \node (5) at (6,0) [circle,draw,minimum size=4mm,inner sep=0pt,fill=red!70, label=above:{$r^2$}] {l};
    \node (6) at (7,1.3) [circle,draw,minimum size=4mm,inner sep=0pt, label=above:{0}]{$\lambda'$};
    \node (7) at (9,1.3) [circle,draw,minimum size=4mm,inner sep=0pt,fill=red!70, label=above:{$\frac{-r^3}{\sqrt 2}$}] {$\rho'$};
    \node (6p) at (7,-1.3)  [circle,draw,minimum size=4mm,inner sep=0pt, label=below:{0}]{$\lambda$};
    \node (7p) at (9,-1.3) [circle,draw,minimum size=4mm,inner sep=0pt,fill=red!70, label=below:{$\frac{-r^3}{\sqrt 2}$}] {$\rho$};
    \node (8) at (10,0)  [circle,draw,minimum size=4mm,inner sep=0pt, label=above:{0}]{r};   
    \node (9) at (12,0)  [circle,draw,minimum size=4mm,inner sep=0pt,  fill=red!70,label=above:{$r^4$}]{l};

    \draw (1) -- (2) node[pos=0.2, above] {$t_1$};
    \draw (1) -- (2p) ;
    \draw (3) -- (4) ;
    \draw (3p) -- (4)  ;
    \draw (5) -- (6) ;
    \draw (5) -- (6p) ;
    \draw (7) -- (8) ;
    \draw (7p) -- (8) ;

    \draw[line width=2.2pt, green!50!black] (2) -- (3) node[midway, above, text=black] {$t_2$};

     \draw[line width=2.2pt, green!50!black] (2p)--(3p);
     \draw[line width=2.2pt, green!50!black] (4)--(5);
     \draw[line width=2.2pt, green!50!black] (6)--(7);
     \draw[line width=2.2pt, green!50!black] (6p)--(7p);
     \draw[line width=2.2pt, green!50!black] (8)--(9);
       
    \draw[dotted , line width=1.2pt] (9) -- ++(1,1);
    \draw[dotted , line width=1.2pt] (9) -- ++(1,-1);


\end{tikzpicture}
\caption{{A semi-infinite chain. The edge state at $E=0$ is localized on the red sites
and the relative amplitudes of the wavefunction are indicated. The ratio 
$r$ is defined as $r=\frac{\sqrt{2}t_1}{t_2}$.}}
\label{fign}
\end{figure}

{The winding number $\nu = 1$ is directly related to the existence of edge states under open boundary conditions.
This can be understood intuitively by considering a semi-infinite chain shown in Fig.\,\ref{fign}. 
We look for an edge state at $E=0$, localized on the sites $nl, n\rho, n\rho'$, 
shown in red in the figure (chirality $\Gamma=-1$), and symmetric under the action of the parity operator ($P=1$). }

{We now determine the wavefunction and impose the condition of localization at the left edge. 
The Schrödinger equation Eq.\,\ref{seq}, projected onto the nodes $1\lambda'$ and $1r$ reads:
\begin{eqnarray}
&&E\Psi(1,\lambda')=t_1\Psi(1,l)+t_2\Psi(1,\rho'),\\
&&E\Psi(1,r)=t_2\Psi(2,l)+t_1\Psi(1,\rho)+t_1\Psi(1,\rho').
\end{eqnarray}
Solving this system for $E=0$ and using the symmetry condition 
$\Psi(1,\rho)=\Psi(1,\rho')$, we obtain
\begin{eqnarray}
\Psi(2,l)=\frac{2t_1^2}{t_2^2}\Psi(1,l)=r^2\Psi(1,l).
\end{eqnarray}
}

{This result generalizes to the following propagation law for the wavefunction:
\begin{eqnarray}\label{edge1n}
\Psi(n+1,l)=\left(\frac{2t_1^2}{t_2^2}\right)^n \Psi(1,l).
\end{eqnarray}
}

{In order to obtain an edge state, the wavefunction must decay into the bulk, i.e., 
its amplitude must vanish as $n \to \infty$. This requires:
\begin{eqnarray}
\frac{2t_1^2}{t_2^2}<1.
\end{eqnarray}
This condition is satisfied precisely in the topological phase with winding number $\nu=1$ from Eq.\,26
where $r^2=\frac{2t_1^2}{t_2^2}$.}

{In the hexagonal chain, tunneling from site $1l$ to $1r$ can occur via two distinct 
paths along the two arms of the hexagon. This introduces an additional degree of 
freedom compared to the standard SSH model, which can be recovered by removing 
the sites $n\lambda$ and $n\rho$.
In this case, the propagation coefficient of the edge state is reduced such that one obtains
\begin{eqnarray}
\Psi(2,l)^{SSH}=\frac{t_1^2}{t_2^2}\Psi(1,l)^{SSH},
\end{eqnarray}
leading to the usual condition for the existence of edge states in the SSH model,
\begin{eqnarray}
\frac{t_1^2}{t_2^2}<1.
\end{eqnarray}
}

{The bulk–boundary correspondence principle applies also to finite systems with two edges. 
In a topological insulating phase with winding number $\nu=1$, 
two edge states are expected, localized near each end of the chain and with energies very close to zero. 
This behavior is analogous to that of the SSH model, but differs from other multi-site unit-cell systems, 
where edge states may appear at finite energies, as in trimer chains \cite{alvarez2019, anastasiadis2022, verma2024}. 
Motivated by this, we examine how these states emerge in an open-boundary system.}

\subsection{Quantum States for the open finite chain}

In the following sections, we measure energy in units of $t_2$ and use the parameter $r$ from Eq.\,\ref{ratio}
to distinguish between different regimes of the system.

Our goal is to determine the eigenstates of the finite chain described by the Hamiltonian in Eq.\,(\ref{hlant}),
by imposing vanishing boundary conditions on its eigenvectors, denoted by $\Phi$. 
Specifically, we require that the projections of $\Phi$ vanish on the boundary sites $n=0, r$ and $n=N+1, l$:
\begin{eqnarray}\label{edge2}
	\Phi(0,r)=0, \\ 
	\label{edge3}
	\Phi(N+1,l)=0. 
\end{eqnarray}

We begin by considering an infinite chain ($N \to \infty$), with eigenstates parametrized by $k$ and $\phi_k$:
$E(k,\phi_k)\equiv E_k^{-}, \Psi(k,\phi_k)\equiv \Psi_k^{-}$.
Here, the wave number $k$ is a continuous variable, while the angle $\phi_k$ is implicitly dependent on $k$.
From Eq.\,\ref{ek}, we invoke the symmetry:
\begin{equation}
E(k,\phi_k)=E(-k,-\phi_k),
\end{equation}
which implies that any linear combination of the form:
\begin{eqnarray}\label{edge1v}
\Phi_k=\alpha \Psi(k,\phi_k)+\beta \Psi(-k,-\phi_k), 
\end{eqnarray}
is also an eigenstate with the same energy:
\begin{eqnarray}
	H|\Phi_k \rangle= E(k,\phi_k) |\Phi_k \rangle.
\end{eqnarray}

We now seek specific combinations as in Eq.\,\ref{edge1v} that also satisfy the boundary conditions from Eqs.\,\ref{edge2} and \ref{edge3}.
Substituting Eqs.\,\ref{psil} and \ref{psir} into Eq.\,\ref{edge1v} yields:
\begin{eqnarray}\label{ed1}
	\Phi_k(nr)&&=\alpha e^{ikn} + \beta e^{-ikn}, \\ 
	\label{ed2}
	\Phi_k(nl)&&=\alpha e^{ikn-i\phi_k} + \beta e^{-ikn+i\phi_k}. 
\end{eqnarray}
From Eqs.\,\ref{ed1} and \ref{edge2}, we find $\Phi_k(0,r)=\alpha+\beta=0$, which implies $\alpha=-\beta$.
Substituting this into Eq.\,\ref{ed2} gives $\Phi_k(n,l) = 2 i \alpha \sin(kn - \phi_k)$.
Imposing the second boundary condition from Eq.\,\ref{edge3} leads to the constraint
$\sin[k(N+1) - \phi_k] = 0$,
implying the angle values $\phi_k = k(N+1) \mod \pi$.

We first consider the solution:
\begin{equation}\label{unghi1}
\phi_k =k(N+1), 
\end{equation}	
and insert it into Eq.\,\ref{edge1v} with $\alpha = -\beta$.
Using the components from Eqs.\,\ref{psil}-\ref{psiro} of $\Psi_k^{-}$,
we obtain the following eigenvector for the finite chain:
\begin{eqnarray}\label{ve}
	|\Phi_k\rangle=\frac{1}{C_k}\sum_{n=1}^N\sum_{s}\Phi_k(ns)|ns\rangle, 
\end{eqnarray} 
with components $\Phi_k(n,s)$ being sinusoidal functions,
\begin{eqnarray}
	\label{phil0}
	\Phi_{k}(nl)&&=\sin k(n-N-1),   \\ 
	\label{phir0}
	\Phi_{k}(nr)&&= \sin k n,  \\ 
	\label{philm0}
	\Phi_{k}(n \lambda/\lambda')&&= \frac{-1}{\sqrt 2}  \sin k \left(n-\frac{1}{2}\right), \\ 
	\label{phiro0}
	\Phi_{k}(n \rho/\rho')&&= \frac{-1}{\sqrt 2}  \sin k \left(n-N-\frac{1}{2}\right). 
\end{eqnarray}
The normalization constant is given by $C_k^2=2N-\frac{\sin kN \cos k(N+\frac{1}{2})} {\sin\frac{k}{2}}$.
The corresponding energy is obtained by substituting the angle from Eq.\,\ref{unghi1} 
into Eq.\,\ref{ek}:
\begin{eqnarray}\label{ek3}
	E_k =e^{ikN}- r e^{ikN+\frac{ik}{2}}. 
\end{eqnarray}


\begin{figure}[h]
\centering
\includegraphics[width=\columnwidth, trim=0 25 10 90]{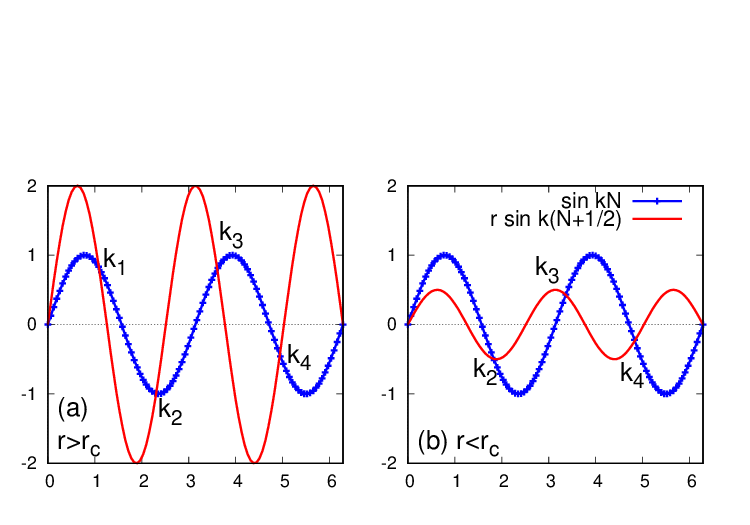}
\caption{The two functions, $\sin kN$ and $r \sin k(N+\frac{1}{2})$, versus $k$, for
		$N=2$ and $r=2$ (a) and $r=0.5$ (b). In (a), Eq.\,\ref{keq2} has $2N=4$ 
		solutions in the interval $(0,2\pi)$,
		denoted $k_1,\cdots,k_4$.
		In (b), the solution $k_1$ is missing.}
\label{functiek}
\end{figure}

For the open boundary chain, not all values of $k$ correspond to valid eigenstates. 
Only those wave numbers that yield real energies in Eq.\,\ref{ek3} are allowed.
To determine them, we extract the quantization condition from the imaginary part of Eq.\,\ref{ek3}:
\begin{equation}\label{keq2}
	\sin kN = r \sin k\left(N+\frac{1}{2}\right). 
\end{equation}
In Fig.\,\ref{functiek}(a), the above involved functions are plotted for $N=2$ and $r=2$, 
finding 4 solutions in the interval $(0,2\pi)$: $k_1, k_2, k_3, k_4$.
In general, this equation admits $2N$ solutions $k_1,\cdots, k_{2N}$ in the interval $k\in (0,2\pi)$ only if:
\begin{equation}\label{cond1}
	r>\frac{N}{N+\frac{1}{2}}. 
\end{equation}
Otherwise, if $r$ becomes smaller than the critical value, the first solution $k_1$ is missing, as shown in Fig.\ref{functiek}(b).
In the limit $N\to \infty$, this threshold $r_c=\frac{N}{N+\frac{1}{2}}$ becomes $r_c=1$, matching the periodic boundary case.

Assuming Eq.\,\ref{cond1} is satisfied, the $2N$ allowed $k$ values satisfy:
\begin{equation}\label{keq3}
	k_\nu\in \left[(\nu-1)\frac{\pi}{N},\nu\frac{2\pi}{2N+1}\right], \quad \nu=1, \cdots, 2N. 
\end{equation}
This property is useful in determining the sign of the energy.
From Eq.\,\ref{ek3}, eliminating $r$, we obtain:
\begin{equation}\label{ek4}
	E_{k_\nu}=\frac{  \sin \frac{k_\nu}{2} } { \sin k_\nu \left(N+\frac{1}{2}\right) }. 
\end{equation}
By substituting Eq.\,\ref{keq3} into Eq.\,\ref{ek4},
one finds that odd $\nu$ indices correspond to positive energies, and even $\nu$ to negative energies. 
Alternatively, the energy can be expressed as:
\begin{equation}\label{ek1r}
	E_{k_\nu}=(-1)^{\nu+1} \sqrt{1+r^2-2r\cos \frac{k_\nu}{2}}, 
\end{equation}
which also preserves the energy sign information.

The next choice for the angle $\phi_k$ is:
\begin{eqnarray}
\phi_k = k(N+1)+\pi.
\end{eqnarray}
This reverses the sign of both $E_k$ in Eq.\,\ref{ek1r}
and the components $\Phi_k(n,l), \Phi_k(n,\rho), \Phi_k(n,\rho')$
from Eqs.\,\ref{phil0} and \ref{phiro0},
in agreement with the chiral symmetry (\ref{ch2}).

The degenerate flat-band eigenstates from Eqs.\,\ref{fb1}, \ref{fb2} and \ref{fb3} also satisfy the Schrödinger equation 
for the open system, since they
contain no $r$ or $l$ components, and thus automatically satisfy the boundary conditions (\ref{edge2}) and (\ref{edge3}).
Therefore, the flat-band energies in Eq.\,\ref{eflat} remain part of the spectrum of the open chain.

\begin{figure}[h]
	\centering
	\includegraphics[width=0.9\columnwidth, trim=10 30 0 20]{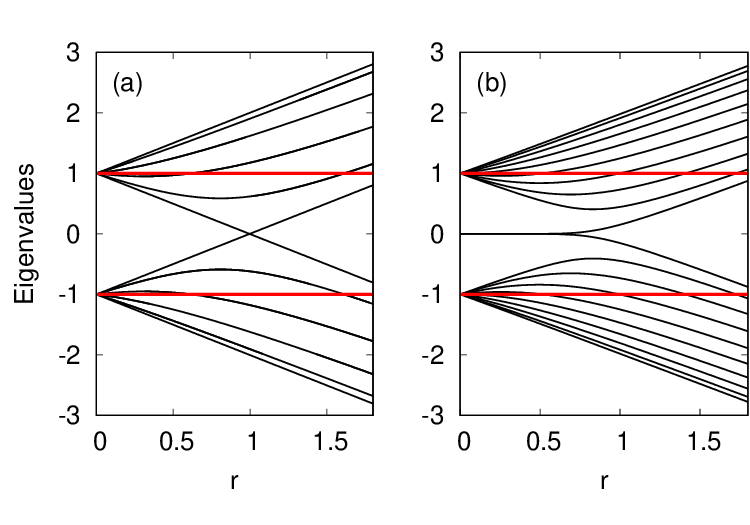}
	\caption{
	Eigenvalues as a function of the hopping ratio $r=\sqrt{2}t_1/t_2$,
	with energy in units of $t_2$. The spectrum under periodic boundary conditions is shown in (a), 
	and under open boundary conditions in (b).
	For each $r$, the flat band exhibits a degeneracy $g=N$. The chain length is $N=5$.
	}
	\label{fsp}
\end{figure}

\subsection{Edge state solution}

We now focus on the parameter range of the hopping ratio $r$ in Eq.\,\ref{ratio} in the range
\begin{equation}\label{cond2}
	r < \frac{N}{N+\frac{1}{2}}. 
\end{equation}
In the extreme limit $r \to 0$ (equivalently, $t_1 \to 0$), 
the edge sites $(1, l)$ and $(N, r)$ shown in Fig.\,\ref{lant1} become effectively isolated, 
which leads to the emergence of two localized zero-energy edge states. 
Therefore, we expect this type of solution to appear for $r < r_c$.
In this case, Eq.\,\ref{keq2} admits only $2N - 1$ real solutions for the wave number $k$.
This is illustrated in Fig.\,\ref{functiek}(b) for $N = 2$, where the solution $k_1$ is missing. 
{In this circumstance, it is natural to take this wave number as imaginary, 
representing an analytical continuation into the complex plane:}
\begin{eqnarray}\label{k1im}
	k_1 = i\delta.
\end{eqnarray}

{The imaginary valuea of k
may also have a different physical interpretation.
For open boundary conditions, translational invariance is lost and the eigenstates are no longer Bloch waves. 
However, bulk-like solutions can still be formed as quantum superpositions of plane waves with real wave vector 
k, as shown in Eq.~\ref{edge1v}. Edge states, in contrast, remain confined near the boundaries and do not preserve translational invariance. 
This is seen, for instance, in the semi-infinite chain (see Eq.~\ref{edge1n}), where the exponential decay from the boundary 
can be obtained by taking a complex value of the wave number $k=i\delta$.}

By substituting Eq.\,\ref{k1im} into Eq.\,\ref{keq2} one obtains:
\begin{equation}\label{deltaeq}
	\sinh \delta N = r \sinh \delta \left( N + \frac{1}{2} \right),
\end{equation}
which determines the solution of $\delta$. 

By using the wave number $k_1 = i\delta$ in Eq.\,\ref{ek3} and its complex conjugate, two expressions for the energy
are obtained,
\begin{equation}
	E_\delta = e^{-\delta N} \left(1 - r e^{-\delta/2}\right) = e^{\delta N} \left(1 - r e^{\delta/2}\right),
\end{equation}
which lead to:
\begin{equation}\label{endelta}
	E_\delta = \sqrt{1 + r^2 - 2r \cosh \frac{\delta}{2}}.
\end{equation}
The corresponding eigenvector is obtained by substituting $k_1 = i\delta$ 
into the vector $|\Phi_k\rangle$ defined in Eq.\,\ref{ve},
and using the imaginary angle $\phi_{k_1}$ from Eq.\,\ref{unghi1}.
This yields:
\begin{equation}\label{ve2}
	|\Phi_\delta \rangle = \frac{1}{C_\delta} \sum_{n,s} \Phi_\delta(ns) |ns\rangle, 
\end{equation}
with components from Eqs.\,\ref{phil0}-\ref{phiro0} replaced by:
\begin{eqnarray}
	\label{phil1}
	\Phi_\delta (nl) &&= \sinh \delta \left(n - N - 1\right),   \\ 
	\label{phir1}
	\Phi_\delta (nr) &&= \sinh \delta n,  \\ 
	\label{philm1}
	\Phi_\delta (n \lambda/\lambda') &&= \frac{-1}{\sqrt{2}} \sinh \delta \left(n - \frac{1}{2}\right), 
	\\
	\label{phiro1}
	\Phi_\delta (n \rho/\rho') &&= \frac{-1}{\sqrt{2}} \sinh \delta \left(n - N - \frac{1}{2}\right). 
\end{eqnarray}
The normalization constant is
$C_\delta^2 = -2N + \frac{\sinh \delta N \cosh \delta \left(N + \frac{1}{2}\right)}{\sinh \frac{\delta}{2}}$.

The resulting eigenvector is localized at the edges of the hexagonal chain, 
with localization length on the order of $1/\delta$ for large $N$.
Since all quantum states come in pairs, 
a corresponding negative-energy eigenvector is obtained by applying the chirality operator $\Gamma$ defined in Eq.\,\ref{ch1}.
These two eigenvectors, $|\Phi_\delta \rangle$ and $\Gamma |\Phi_\delta \rangle$, show that when open boundary conditions are imposed,  
two topological edge states appear inside the energy gap of the insulating phase — but only for the parameter range $r < r_c$.  
Such edge states are absent in the trivial insulating phase where $r > r_c$.

In Fig.\,\ref{fsp} we show the eigenvalues of the finite chain and the emergence of mid-spectrum edge states.
Figure~\ref{fsp}(a) displays the energy spectrum as a function of the hopping parameters ratio $r$ 
for a chain of length $N = 5$ with periodic (cyclic) boundary conditions. 
A clear energy gap is observed, which closes at the critical point $r_c = 1$,
signaling a topological phase transition between a topological insulator ($r < 1$) and a trivial one ($r > 1$).
When open boundary conditions are applied, Fig.\,\ref{fsp}(b) reveals the appearance of two edge states 
in the topological phase ($r < r_c$).
In agreement with the bulk-boundary correspondence principle, this phase is characterized by a non-zero winding number
and this is shown in the Appendix\,B.
Note that in the open chain case, the critical value $r_c$ becomes size-dependent,
varying with the system length $N$.

\begin{figure}[h]
	\centering
	\includegraphics[width=0.52\columnwidth, trim=20 70 30 70]{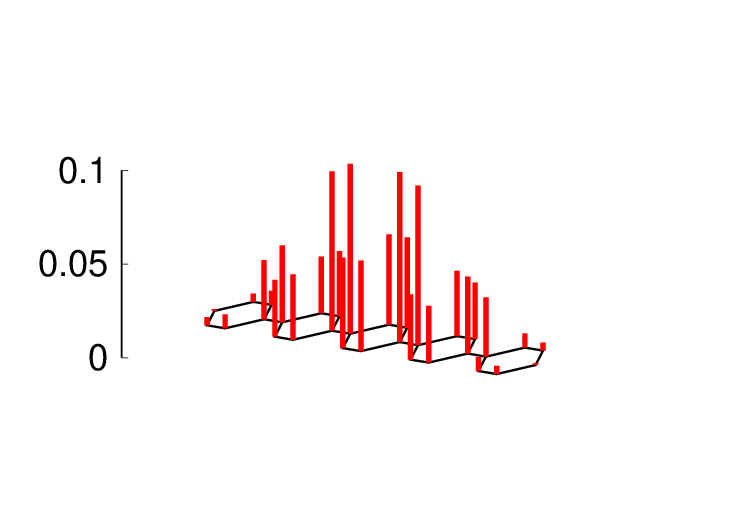}
	\includegraphics[width=0.43\columnwidth, trim=50 70 50 70]{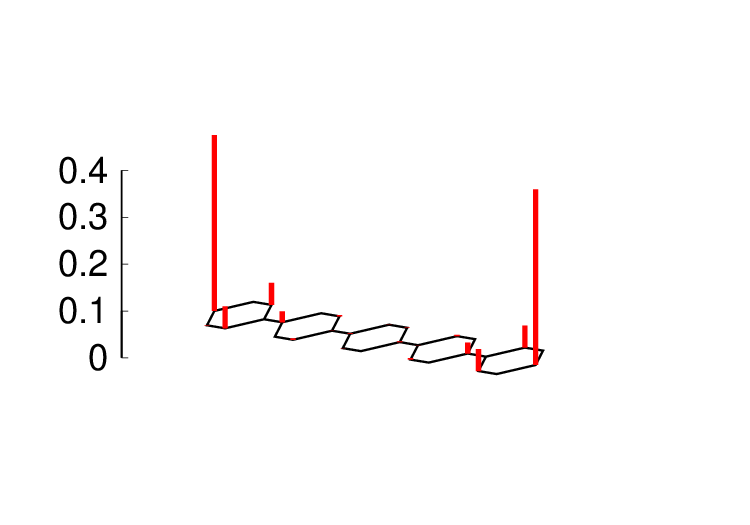}
	\caption{
	Two typical eigenvectors, a bulk $|\Phi_{k_\nu}(n,s)|^2$ (left) 
	and an edge state $|\Phi_{\delta}(n,s)|^2$ (right) for a chain with open boundary conditions in the topological phase. 
	The chain length is $N=5$ and the hopping ratio is $r=0.5$ with $r<r_c$. 
	}
	\label{vec}
\end{figure}

Two relevant eigenvectors for the open boundary chain are shown in Fig.\,\ref{vec}.
For $N=5$ the critical ratio is $r=0.91$ so one chooses $r=0.5$ such that the system is in the 
topological phase (exhibiting both bulk and edge states).
The wave numbers equation Eq.\,\ref{ek4} has $2N-1$ solutions, $k_2,\cdots,k_{10}$. 
The wave function $|\Phi_{k_\nu}(n,s)|^2$ from Eqs.\,\ref{phil0}-\ref{phiro0}
for $\nu=10$ is shown in the left of the figure.
It has the wave number 
$k_{10}=5.67$ and energy from Eq.\,\ref{ek1r} is $E_{k_{10}}=-1.48t_2$.
The imaginary wave number $k_1=i\delta$ is calculating by using the Eq.\,\ref{deltaeq} 
which gives $\delta=1.386$.
In this case one has two middle spectrum edge states $\Phi_{\delta}$ and $\Gamma\Phi_{\delta}$.
The squared wave function $|\Phi_{\delta}(n,s)|^2$ calculated from Eqs.\,\ref{phil1}-\ref{phiro1} 
is depicted in the right of Fig.\,\ref{vec}.
Its energy from Eq.\,\ref{endelta} is $E_{\delta}=0.73 \times 10^{-3}t_2$.

When the hopping ratio is equal to the critical value 
\begin{eqnarray}
r_c=\frac{N}{N+\frac{1}{2}}
\end{eqnarray}
the finite system is at the crossing point from the bulk insulator to the topological insulator.
In this case the wave number of the two middle spectrum states
becomes zero. $k_1=i\delta=0$ in Eqs.\,\ref{ek4} or \ref{deltaeq}.
The sinusoidal wave function $\Phi_{k_1}(n,s)$ from Eq.\,\ref{ve} turn into the hyperbolic function $\Phi_{\delta}(n,s)$ from Eq.\,\ref{ve2}.
The critical point states have energies
\begin{eqnarray}
E_c=\pm(1-r_c)=\pm \frac{1}{2N+1}
\end{eqnarray}
and the correspnding wave functions are obtained from Eqs.\,\ref{phil1}-\ref{phiro1} as a limiting case.


\section{Conclusions}

In this paper we have investigated the eigenstates and topological properties of a molecular wire 
containing hexagon unit cells coupled linearly in the para positions.  
Both dispersive and flat bands are obtained in the energy spectrum.  
By a proper choice of two different hopping energies we are able to identify two insulator regimes  
with a gap closing in between.  They are characterized by their different winding numbers:
$\nu = 1$ in the topological phase and $\nu = 0$ in the trivial phase.

For the linear chain with open boundaries, we have proven the existence of two edge states  
located in the non-trivial insulator gap, while the normal insulator gap persists.  
The critical point of the transition is calculated for both cyclic and vanishing boundary conditions.


\begin{acknowledgements}
We acknowledge financial support from the Core Program of the National Institute of Materials Physics,
granted by the Romanian MCID under projects
 no.\,PC2-PN23080202 and PC4-PN23080404.
 The author used ChatGPT to improve the language and readability.
After using this tool, the author reviewed and edited the content 
and takes the full responsibility for the final version.
\end{acknowledgements}

\appendix

\section{Eigenstates of $H_k$}

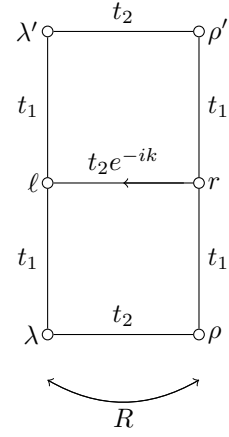
\begin{figure}[h!]
    \centering
\begin{tikzpicture}[scale=2, every node/.style={font=\normalsize}]
\node (a) at (0,0) [circle,draw,minimum size=4pt,inner sep=0pt] {}; 
\node (b) at (0,1) [circle,draw,minimum size=4pt,inner sep=0pt] {}; 
\node (c) at (0,2) [circle,draw,minimum size=4pt,inner sep=0pt] {}; 

\node (d) at (1,0) [circle,draw,minimum size=4pt,inner sep=0pt] {}; 
\node (e) at (1,1) [circle,draw,minimum size=4pt,inner sep=0pt] {}; 
\node (f) at (1,2) [circle,draw,minimum size=4pt,inner sep=0pt] {}; 

\draw (a) -- (b) node[midway,left] {$t_1$};
\draw (b) -- (c) node[midway,left] {$t_1$};
\draw (d) -- (e) node[midway,right] {$t_1$};
\draw (e) -- (f) node[midway,right] {$t_1$};

\draw (a) -- (d) node[midway,above] {$t_2$};
\draw (b) -- (e) node[midway,above] {$t_2 e^{-ik}$};
\draw (c) -- (f) node[midway,above] {$t_2$};

\node[left]  at (b) {$\ell$};
\node[left]  at (c) {$\lambda'$};
\node[right] at (d) {$\rho$};
\node[right] at (f) {$\rho'$};
\node[left]  at (a) {$\lambda$};
\node[right]  at (e) {$r$};

\draw[->,bend right] (0,-0.3) to (1,-0.3);
\draw[->,bend left] (1,-0.3) to (0,-0.3);
\node at (0.5,-0.55) {$R$};

\draw[<-] (0.5,1) -- (0.9,1);
\end{tikzpicture}
 \caption{The graphic reprezentation of $H_k$ Hamiltonian.}
    \label{graphhk}
\end{figure}

In Fig.\,\ref{graphhk} we illustrate the Hamiltonian graph of $H_k$. 
The graph nodes represent the $k$-basis vectors $|ks\rangle$ with $s=l,r,\lambda, \lambda', \rho$ and $\rho'$. 
The graph edges correspond to the non-zero matrix elements of $H_k$, with their values indicated in the figure. 
For instance, $\langle kl |H_k | kr \rangle = t_2 e^{-ik}$ is shown as an oriented edge.

We observe that under the graph reflection with respect to the vertical axis, 
as indicated by the operation $R$ in the figure, 
the Hamiltonian becomes complex conjugated, since $t_2 e^{-ik}$ changes to $t_2 e^{ik}$. 
This implies that
\begin{eqnarray}\label{sstar}
H_k^{*}=R H_k R^{+},
\end{eqnarray}
where the $k$-subspace parity operation $R$ is defined as
\begin{eqnarray}
R=|k\lambda\rangle \langle k \rho | 
+ |k\lambda'\rangle \langle k \rho' | 
+ |kl\rangle \langle k r | + h.c.,
\end{eqnarray}
with $R=R^{+}=R^{-1}$.

Let us consider an eigenvector $|\Psi_k \rangle$ of $H_k$ from Eq.\,\ref{hkek}, corresponding to the energy $E_k$. 
Due to the symmetry expressed in Eq.\,\ref{sstar}, 
the mirror vector $R|\Psi_k\rangle$ is also an eigenvector of $H_k^*$. 
This means that the two vectors $R|\Psi_k\rangle$ and $|\Psi_k^*\rangle$ 
must be equal up to a phase factor $e^{i\phi_k}$:
\begin{eqnarray}\label{sr}
|\Psi_k^*\rangle = e^{i\phi_k} R|\Psi_k\rangle.
\end{eqnarray}

For instance, by projecting the above equation on the left with the vectors 
$\langle kl|$, $\langle k\lambda|$, and $\langle k\lambda'|$, 
we obtain the following relations between the components of the eigenvectors:
\begin{eqnarray}\label{xls}
\Psi_{kl}^*&&=e^{i\phi_k} \Psi_{kr},\\
\label{xlms}
\Psi_{k\lambda}^*&&=e^{i\phi_k} \Psi_{k\rho},\\
\label{xros}
\Psi_{k\lambda'}^*&&=e^{i\phi_k} \Psi_{k\rho'}.
\end{eqnarray}

We now aim to determine the symmetric eigenvectors that solve Eq.\,\ref{hkek}. 
First, they must have parity $P=1$ of the operator defined in Eq.\,\ref{par}, i.e.,
\begin{eqnarray}
\Psi_{k\lambda}&&=\Psi_{k\lambda'},\\
\label{xrho}
\Psi_{k\rho}&&=\Psi_{k\rho'}.
\end{eqnarray}

Next, we write the equations obtained by projecting Eq.\,\ref{hkek} on the left with the vectors 
$\langle kr|$ and $\langle k\lambda|$, taking into account the explicit form of the Hamiltonian in Eq.\,\ref{hka} 
or, equivalently, using the graph in Fig.\,\ref{graphhk}:
\begin{eqnarray}
E_k \Psi_{kr}&&=t_2 e^{ik} \Psi_{kl}+t_1 \Psi_{k\rho} + t_1 \Psi_{k\rho'},  \\
E_k \Psi_{k\lambda}&&= t_1 \Psi_{kl} + t_2 \Psi_{k\rho}.
\end{eqnarray}

We now substitute into the above equations the relations \ref{xls}, \ref{xlms}, 
and the equality \ref{xrho}. 
Since the normalization of the wavefunction will be imposed at the end, 
for simplicity we may set $\Psi_{kr}=1$. 
From Eq.\,\ref{xls} this implies $\Psi_{kl}=e^{-i\phi_k}$. 
Finally, we obtain the equations
\begin{eqnarray}
&&E_k = t_2 e^{ik-i\phi_k}+2 t_1 e^{-i\phi_k}\Psi_{k\lambda}^*~~\text{and}\\
&&E_k \Psi_{k\lambda}= t_1 e^{-i\phi_k} + t_2 e^{-i\phi_k}\Psi_{k\lambda}^*,
\end{eqnarray}
from which the expressions for $E_k$ and $\Psi_{k\lambda}$ are obtained.
To perform this, we rewrite these two equations for $t_1=0$ with arbitrary $t_2$, 
and then for $t_2=0$ with arbitrary $t_1$, 
and we obtain the vector component:
\begin{eqnarray}
\Psi_{k\lambda}=\pm \frac{1}{\sqrt 2} e^{-i\frac{k}{2}}.
\end{eqnarray}
The component $\Psi_{k\rho}$ is obtained from Eq.\,\ref{xlms}. 
The final eigenvector is written in Eq.\,\ref{vk}, and its corresponding energy in Eq.\,\ref{ek}.

\section{The critical point state vector}

When the ratio $r_c=\frac{\sqrt{2}t_1}{t_2}=\frac{N}{N+1/2}$ the wave number becomes zero, $k_1=i\delta=0$.
For the eigenvalue
\begin{eqnarray}
E_c=1-r_c=\frac{1}{2N+1}
\end{eqnarray}
the correspnding wave functions are obtained from Eqs.\,\ref{phil1}-\ref{phiro1} as a limiting case.
One has
\begin{equation}\label{vcr}
	|\Phi_c \rangle = \frac{1}{C} \sum_{n,s} \Phi_c(ns) |ns\rangle, 
\end{equation}
with the following components: 

\begin{eqnarray}
	\label{philcr}
	\Phi_c (nl) &&= n - N - 1,   \\ 
	\label{phircr}
	\Phi_c (nr) &&= n,  \\ 
	\label{philmcr}
	\Phi_c (n \lambda/\lambda') &&= \frac{-1}{\sqrt{2}} \left(n - \frac{1}{2}\right), 
	\\
	\label{phirocr}
	\Phi_c (n \rho/\rho') &&= \frac{-1}{\sqrt{2}}  \left(n - N - \frac{1}{2}\right). 
\end{eqnarray}

The normalization constant is
$C^2 = N\frac{8N^2+6N+1}{6}$.

\end{document}